\documentstyle[12pt]{article} 

\newcommand{\lsim}{\mbox{\,\raisebox{0.13cm}{$<$} \hspace{-1.45em} 
            \raisebox{-0.13cm}{$\sim$}\,}}

\newcommand{\pslash}{\mbox{$\not{\hspace{-0.8mm}p}$}}

\newcommand{\qslash}{\mbox{$\not{\hspace{-0.8mm}q}$}}
\newcommand{\Dslash}{\mbox{$\not{\hspace{-1.1mm}D}$}}
\newcommand{\vslash}{\mbox{$\not{\hspace{-0.8mm}v}$}}

\newcommand{\kslash}{\mbox{$\not{\hspace{-0.8mm}k}$}}
\newcommand{\lslash}{\mbox{$\not{\hspace{-0.3mm}l}$}}

\newcommand{\hoch}[1]{\mbox{\rule[0cm]{0cm}{#1}}}
\newcommand{\nc}{\newcommand}

\nc{\be}{\begin{equation}}
\nc{\ee}{\end{equation}}
\nc{\bea}{\begin{eqnarray}}
\nc{\eea}{\end{eqnarray}}
\nc{\dn}{\Delta _0}
  
\begin{document}
\begin{titlepage}
 \begin{flushleft}
   MZ-TH/93-32\\ 
   November 1993
 \end{flushleft}
 \begin{center}
 {\Large Inclusive Semileptonic $B$ Decays in QCD
 including Lepton Mass Effects}\\[2cm]
\large
 S. Balk$^1$ \\  J.G. K\"orner$^2$ \\  D. Pirjol$^1$ \\
 and \\ K. Schilcher $^2$ \\[.5cm]
 Johannes Gutenberg-Universit\"at\\
 Institut f\"ur Physik, Staudingerweg 7\\
 D-55099 Mainz, Germany\\[2cm]
\normalsize
 \begin{abstract}  
 Starting from an Operator Product Expansion in the Heavy 
 Quark Effective Theory up to order $1/m_b^2$ we calculate the
 inclusive semileptonic decays of unpolarized bottom hadrons 
 including lepton mass effects. We calculate 
 the differential decay spectra $d\Gamma/(dE_\tau )$,
 and the total decay rate for B meson decays to final states containing
 a $\tau$ lepton.
\end{abstract}  
\end{center}

$\quad $\\
\footnotesize
$^1\,$Supported by the Graduiertenkolleg Teilchenphysik, Universit\"at Mainz\\
$^2\,$Supported in part by the BMFT, FRG under contract 06MZ730\\
\normalsize
\end{titlepage}
\section{Introduction}
Recently, the theoretical investigation of inclusive semileptonic bottom 
hadron ($H_b$) decays has witnessed a renaissance through the use of the 
Operator Product Expansion (OPE) in the Heavy Quark Effective Theory (HQET) 
\cite{chay,mannel}.
 Up to now these investigations have been limited to the 
discussion of the zero lepton mass case. While this is a good 
approximation for inclusive semileptonic $H_b$ decays involving an electron 
or a muon, a study of inclusive semileptonic $H_b$ decays involving a 
$\tau $ must certainly include the discussion of lepton mass effects. 

There are always two aspects to the lepton mass problem in semileptonic 
decays. First the decay dynamics changes as one is probing also the spin 
zero components of the exchanged gauge boson in addition to the spin one 
components active in the zero lepton mass case. In the formalism of 
semileptonic inclusive decays this means that there are two more structure 
functions ($W_4, W_5$) 
contributing to the decay as compared to the zero lepton mass  
case ($W_1, W_2, W_3$).
 In addition the decay kinematics gets changed and becomes 
somewhat more complicated than in the zero lepton
 mass case. As the very dynamics of the 
OPE method entangles the decay kinematics of 
particle and parton decay in a subtle manner one needs 
to carefully discuss the decay kinematics of the massive lepton case.

Apart from theoretical considerations the calculation of inclusive 
semileptonic bottom decays has immediate phenomenological implications 
since current high energy experiments have reached a stage of 
sophistication that allow them to start measuring inclusive semileptonic 
$H_b$ rates involving the $\tau$. For example, 
recently the ALEPH collaboration reported on the first measurement 
of the inclusive semileptonic bottom hadron decay $H_b\to X\tau^-\bar 
\nu_{\tau}$ \cite{ale}. They quote a branching ratio 
\bea
 Br(H_b\to X\tau\bar\nu_\tau) &=& (2.76\pm 0.47\pm 0.43)\% \mbox{.}
\eea

In this presentation we shall closely follow the notation of \cite{man}, 
albeit with the necessary extensions to include lepton mass effects. We also 
include a separate section on the operator product method in HQET where we 
present general formulae which allow one to go beyond the $1/m_b^2$ results 
considered in \cite{bigi,mannel}.

\section{Operator Product Expansion in HQET}
Inclusive semi-leptonic B-decays are related to the transition operator
\bea
 t_{\mu\nu}&=&-i\int d^4x e^{-iqx} T \bar b(x) \Gamma _{\mu}
   \stackrel{\vert \hspace*{-0.5mm}\vspace*{0mm}
           \overline{\hoch{1.5mm}\hspace*{1cm}}\hspace*{-0.5mm}\vspace*{0mm}
           \vert}{q(x)\bar q(0)}
    \Gamma _{\nu} b(0), \label{eq:trans}
\eea 
where $q(x)$ stands for up or charm quarks. The propagator term in 
 (\ref{eq:trans}) can be written 
in momentum space as
\bea
   \stackrel{\vert \hspace*{-0.5mm}\vspace*{0mm}
           \overline{\hoch{1.5mm}\hspace*{1cm}}\hspace*{-0.5mm}\vspace*{0mm}
           \vert}{q(x)\bar q(0)} &=&
  i S(x,0)=i\int \frac{d^4l}{(2\pi)^4}e^{-ilx} S(l)\mbox{.}
\eea 
The up or charm quarks may be considered 
as propagating in the external gluon field $F_{\alpha\beta}$ generated by 
the heavy quark source. 
The expansion of the quark propagator in powers of the external field is 
best derived in the 
Fock-Schwinger gauge following the general discussion in \cite{novi}.
One obtains 
\bea
 S(l)&=&\left[ 
\frac{1}{\lslash-m}+\frac{1}{\lslash-m}\gamma^\alpha 
\frac{1}{\lslash-m}\gamma^\beta \frac{1}{\lslash-m}
\left( \frac{-ig}{2}\right) F_{\alpha\beta} \right.
    \label{eq:propa}
\\
 & &\left. +\frac{g}{3}\left( 
 D_{\beta} F_{\gamma \alpha}
 + D_{\gamma} F_{\beta \alpha} \right)
\frac{1}{\lslash-m}\gamma^\alpha 
\frac{1}{\lslash-m}\gamma^\beta \frac{1}{\lslash-m}
\gamma^\gamma \frac{1}{\lslash-m}
+ \dots  \right], 
\nonumber
\eea 
where $D_{\alpha}=\partial_{\alpha}+igA^a_{\alpha}t^a$.

The OPE can only be meaningfully applied to 
(\ref{eq:trans}) if short distance (large momentum) contributions are 
absorbed in the Wilson coefficients and the long distance (small momentum) 
contributions are absorbed in the matrix elements of the operators.
We must explicitly extract the large mass scale into an exponential, i.e.
pass to the operators in HQET. We write
\bea
 b(x)&=& e^{-im_b v\cdot x}B(x) \mbox{.}
\eea
The field $B(x)$ is related to the usual field $b_v(x)$ of the HQET by
 \cite{djs}
\bea
B(x)&=&\left[ 1+\frac{i\Dslash _\perp}{2m_b}+
       \frac{1}{4m_b^2}\left(v\cdot D \Dslash _\perp 
       -\frac{1}{2} \Dslash _\perp ^2 
        \right)        \right] b_v(x) \mbox{,}
\eea
where $D_\perp ^\mu = D^{\mu} -v^{\mu}v\cdot D$. 
Using the expansion
\bea
 B(x)&=& B(0)+(x\cdot D)B(0)+\frac{1}{2}(x\cdot D)^2 B(0)+\dots
   \nonumber \\
  &=& e^{-ikx}B(0),
\eea
where $k_\mu=iD_\mu$ relates to the residual momentum of the heavy quark, 
we obtain from (\ref{eq:trans})
\bea
 t_{\mu \nu} &=&\int \frac{d^4l}{(2\pi)^4}\int d^4x e^{-ix(q-m_b v+l-k)}
    \cdot \bar B(0) \Gamma _\mu S(l) \Gamma _\nu B(0) \mbox{.}
    \label{eq:transb}
\eea
The integrations in (\ref{eq:transb}) are trivially carried out with the 
result
\bea
t_{\mu \nu} &=& 
B(0)\Gamma _\mu S(m_b v +k -q) \Gamma _\nu B(0)
    \mbox{.} \label{eq:master}
\eea
To obtain explicit results from this master formula we must substitute 
for the propagator in the background gluon field from (\ref{eq:propa})
and expand the resulting expressions in powers of $k$
 using the formula
\bea
\frac{1}{\pslash + \kslash -m}&=&
\frac{1}{\pslash -m}
-\frac{1}{\pslash -m}\kslash \frac{1}{\pslash -m}
\nonumber \\
&&
+\frac{1}{\pslash -m}\kslash \frac{1}{\pslash -m}
\kslash \frac{1}{\pslash -m}-\dots +\dots , 
\eea 
where $p=m_b v -q $.

\section{Inclusive Decays $H_b\to X_{c,u}+ \tau + \bar\nu_\tau $}
Consider a heavy hadron $H_b$ moving with a four-velocity $v$ 
and having a mass $M_{H_b}$ (containing a b quark of mass $m_b$) which
decays weakly into a hadronic system of invariant mass $M_X$ 
(containing a c or u quark of mass $m_j$ ($j$=c,u)) and an off-shell
$W^-$ with momentum $q$. The $W^-$ decays subsequently into a
$\tau $-lepton and an antineutrino $\bar\nu_\tau$.
Previous authors have investigated inclusive semileptonic  $H_b$ decays 
involving the $e$- and $\mu$-lepton where the mass of the final lepton can 
be neglected \cite{bigi}-\cite{mannel}. 
We extend the previous work 
to the case of the tau lepton where the $\tau$-mass ($m_\tau$ =1.777 GeV) 
cannot be neglected. We denote the energies of the antineutrino and tau in
the rest-frame of the initial hadron by $E_\nu$ and $E_\tau$, respectively.
 
   The phase space in the variables $(q^2, E_\tau)$ has the 
form shown in Fig.1 for the inclusive decay $B\to X_c\tau\bar\nu_\tau$
and in Fig.2 for the decay $B\to X_u\tau\bar\nu_\tau$. $E_\tau$ may
take values from $m_\tau$ up to $(M_{H_b}^2-(M_{X_{min}}+m_\tau)^2)/
(2M_{H_b})$ and $q^2$ is bounded by \cite{koe}
\bea
 m_\tau^2 \leq q^2 \leq \frac{M_{H_b}(M_{H_b}^2-M_{X_{min}}^2+
   m_\tau^2-2M_{H_b}E_\tau)(E_\tau+\sqrt{E_\tau^2-m_\tau^2})
   +m_\tau^2M_{X_{min}}^2}{M_{H_b}^2+m_\tau^2-2M_{H_b}E_\tau}.\nonumber\\
\eea
  Here $M_{X_{min}} = m_D$ for $b\to c$ and $= m_\pi$ for $b\to u$.

   The hadronic tensor of the inclusive process under consideration
is defined to be 
\bea
 W^{\mu \nu}_j &=& (2\pi)^3 \sum_X \delta^4(p_{H_b}-q-p_X)
 \langle H_b(v,s)\vert J_j^{\dag \mu} \vert X\rangle 
 \langle X \vert J_j^{\nu} \vert H_b(v,s)\rangle, 
\eea
where $J_j^\mu$ is the hadronic current and $j=u,c$.
The expansion of $W^{\mu \nu}$ in terms of Lorentz invariant structure 
functions reads
\bea
W^{\mu \nu}=-g^{\mu \nu}W_1+v^\mu v^\nu W_2 
            -i\epsilon ^{\mu \nu \rho \sigma} v_\rho q_\sigma W_3
            +q^\mu q^\nu W_4 +(q^\mu v^\nu+ v^\mu q^\nu) W_5,  
\eea
where $q=p_\tau+p_\nu$ and 
$W_1,..,W_5$ are the structure functions to be determined from the 
dynamics. Note that there is no T-odd term proportional to
$(q^\mu v^\nu - v^\mu q^\nu)$ in the fully inclusive decay.

The contraction of the hadronic with the leptonic tensor gives 
the spin averaged differential decay rate
\bea
\frac{d\Gamma}{dq^2 dE_\tau dE_\nu} &=&
\frac{\vert V_{jb}\vert ^2 G_F^2}{2\pi^3}
\left(W_1(q^2-m_\tau^2)+W_2 \left(2 E_\tau E_\nu-\frac{q^2-m_\tau^2}{2}\right)
\right.\nonumber\\
&&              +W_3 (q^2 (E_\tau-E_\nu)-m_\tau^2 (E_\nu+E_\tau))
\nonumber\\
&& \left.
                +W_4 \left(\frac{m_\tau^2}{2} (q^2-m_\tau^2)\right)
                +W_5 (2 m_\tau^2 E_\nu) \right) \mbox{.}
                 \label{eq:decayrate}
\eea
In comparison to the massless lepton case the factors which multiply
$W_1$ to $W_3$ change, and there are additional contributions from the 
structure functions $W_4$ and $W_5$. 

The structure functions $W_1,...,W_5$ are given by the absorptive part of 
the matrix element of the transition operator
\bea
 T^{\mu \nu}&=&
\langle H_b \vert t_{\mu \nu} \vert H_b \rangle \\ 
&=& 
-g^{\mu \nu}T_1 +v^\mu v^\nu T_2 -i
\epsilon ^{\mu \nu \rho \sigma }v_\rho q_\sigma T_3
+q^\mu q^\nu T_4 +(q^\mu v^\nu + v^\mu q^\nu) T_5 ,
\nonumber 
\eea
where $Im T^{\mu \nu}= -\pi W^{\mu \nu}$.
What remains to be done is to determine the amplitudes $T_1,..,T_5$.

First we expand the master formula (\ref{eq:master}) 
of our operator product expansion in powers of 
$k$. As we are interested 
only in operators up to dimension five we can terminate the expansion at order 
$k^2$.  
From the expansion of 
\bea
&&\frac{\bar u \gamma^\mu (1-\gamma _5) (m_b \vslash -\qslash + 
\kslash +m_j)\gamma^\nu (1-\gamma _5) u}
{4 ((m_b v -q+ k)^2- m_j^2 +i\epsilon)},
\eea
in powers of $k$ we obtain the contribution of the $k^0$- terms
to $T^{\mu \nu}$:
\bea
 T^{(0)}_1&=&\frac{m_b-v.q}{2 \dn} \nonumber \\
 T^{(0)}_2&=&\frac{m_b}{\dn} \nonumber \\
 T^{(0)}_3&=&\frac{1}{2 \dn} \nonumber \\
 T^{(0)}_4&=&0 \nonumber \\
 T^{(0)}_5&=&\frac{-1}{2 \dn} \mbox{.}
\eea
The $ T_1^{(i)}, T_2^{(i)}, T_3^{(i)}$ terms were already given in
\cite{man}. We list them here only for completeness.
The contributions of the $k^1$- terms to $T^{\mu \nu}$ read:
\bea
 T^{(1)}_1&=& m_b E_b \left( \frac{1}{2 \dn}-\frac{(m_b-v.q)^2}{\dn ^2}
              \right) + \frac{2 m_b}{3} (K_b+G_b)
            \left( \frac{-1}{2 \dn}+\frac{q^2-(v.q)^2}{\dn ^2}
              \right)
\nonumber \\
 T^{(1)}_2&=& m_b E_b \left( \frac{1}{\dn}-\frac{2m_b(m_b-v.q)}{\dn ^2}
              \right) + \frac{2 m_b}{3} (K_b+G_b)
            \left( \frac{1}{\dn}+\frac{2m_b v.q}{\dn ^2}
              \right)     
\nonumber \\
 T^{(1)}_3&=& - m_b E_b \left(\frac{m_b-v.q}{\dn ^2}
              \right) - \frac{2 m_b}{3} (K_b+G_b)
            \left( \frac{m_b-v.q}{\dn ^2}
              \right)         
\nonumber \\
 T^{(1)}_4&=&  \frac{2 m_b}{3} (K_b+G_b)
              \left( \frac{2}{\dn ^2}
              \right)           
\nonumber \\
 T^{(1)}_5&=&  m_b E_b \left( \frac{m_b-v.q}{\dn ^2}
              \right) + \frac{2 m_b}{3} (K_b+G_b)
            \left( \frac{-m_b-v.q}{\dn ^2}
              \right)               \mbox{,}
\eea
where the spin-energy, kinetic energy and total energy matrix elements 
$G_b, K_b$ and $E_b$, respectively, of the b-quark in the hadron $H_b$
 are defined as in \cite{man}.
The order $k^2$ contributions are
\bea
 T^{(2)}_1&=&-\frac{m_b^2 K_b}{3}(m_b-v.q)
  \left(\frac{4(q^2-(v.q)^2)}{\dn ^3 }-\frac{3}{\dn ^2} \right)
\nonumber \\
 T^{(2)}_2&=&-\frac{2m_b^3 K_b}{3}
  \left(\frac{4(q^2-(v.q)^2)}{\dn ^3 }-\frac{3}{\dn ^2} \right)
 +\frac{4m_b^2 K_b v.q}{3 \dn ^2} 
\nonumber \\
 T^{(2)}_3&=&-\frac{m_b^2 K_b}{3}
  \left(\frac{4(q^2-(v.q)^2)}{\dn ^3 }-\frac{3}{\dn ^2} \right)
 +\frac{2m_b^2 K_b}{3\dn ^2} 
\nonumber \\
 T^{(2)}_4&=&0
\nonumber \\
 T^{(2)}_5&=& \frac{m_b^2 K_b}{3}
  \left(\frac{4(q^2-(v.q)^2)}{\dn ^3 }-\frac{5}{\dn ^2} \right)
 \mbox{.}
\eea
There is an additional gluonic contribution originating from 
the gluon operator in the second 
term of (\ref{eq:propa}), i.e.
\be
 \frac{g}{2\dn ^2}\bar b G^{\alpha \beta} 
 \epsilon_{\alpha \beta \lambda \sigma}(m_b-q)^\lambda
 [g^{\mu \sigma }\gamma ^\nu 
 +g^{\nu \sigma }\gamma ^\mu 
 -g^{\mu \nu }\gamma ^\sigma
 +i\epsilon ^{\mu \sigma \nu \tau} \gamma _\tau ]\frac{1-\gamma _5}{2}b
\ee
 which leads to 
 \bea
 T^{(g)}_1&=&-\frac{m_b^2 G_b}{3\dn ^2}(m_b-v.q)
\nonumber \\
 T^{(g)}_2&=&\frac{2m_b^3 G_b}{3\dn ^2}
\nonumber \\
 T^{(g)}_3&=&-\frac{m_b^2 G_b}{3\dn ^2}
\nonumber \\
 T^{(g)}_4&=&0
\nonumber \\
 T^{(g)}_5&=& -\frac{m_b^2 G_b}{3\dn ^2}
\eea
Summing the various contributions 
 $T_i=\sum_z T^{(z)}_i$, $i\in \{0,1,2,g\}$ we obtain 
\bea
T_1&=&\frac{1}{2\dn}(m_b-v.q)(1+X_b)+\frac{2m_b}{3}(K_b+G_b)
\left( \frac{-1}{2\dn}+ \frac{q^2-(v.q)^2}{\dn^2} \right) \nonumber \\
 && +\frac{m_b E_b}{2\dn} -\frac{m_b^2 G_b}{3\dn^2}(m_b-v.q)\nonumber \\
T_2&=&\frac{m_b}{\dn}(1+X_b)+\frac{2m_b}{3}(K_b+G_b)
     \left( \frac{1}{\dn}+\frac{2m_b v.q}{\dn^2}  \right) \nonumber \\
 && +\frac{m_b E_b}{\dn}+\frac{4m_b^2 K_b v.q}{3\dn^2} 
    +\frac{2 m_b^3 G_b}{3 \dn^2} \nonumber \\
T_3&=&\frac{1}{2\dn}(1+X_b)-\frac{2m_b}{3}(K_b+G_b)
     \frac{m_b-v.q}{\dn^2}+\frac{2 m_b^2 K_b}{3 \dn^2}
    -\frac{m_b^2 G_b}{3 \dn^2} \nonumber \\
T_4&=&\frac{4 m_b}{3 \dn^2}(K_b+G_b)\nonumber \\
T_5&=&\frac{-1}{2\dn}(1+X_b)-\frac{2m_b}{3}(K_b+G_b)
     \frac{2m_b+v.q}{\dn^2} + \frac{m_b^2 G_b}{3 \dn^2} \nonumber
\eea
where
\bea
X_b&=&\frac{-2(m_b-v.q)m_b E_b}{\dn}
     -\frac{8m_b^2 K_b}{3 \dn^2}(q^2-(v.q)^2)
     +\frac{2 m_b^2 K_b}{\dn}
     \nonumber
\eea
Our formulae for $T_1,T_2,T_3$ coincide with the ones given in \cite{man}, 
whereas the $T_4$ and $T_5$ expressions are new.
 
We get the resulting differential decay rate (\ref{eq:decayrate}) by
taking the imaginary part of $T^{\mu \nu}$ and dividing it
  by ($-\pi$). This amounts to the replacements
\bea
\frac{1}{\Delta _0}&\rightarrow &
  \delta ((m_b v-q)^2-m_j^2),
\nonumber \\
\frac{1}{\Delta _0^2}&\rightarrow &
  -\delta '((m_b v-q)^2-m_j^2),
\nonumber \\
\frac{1}{\Delta _0^3}&\rightarrow &
\frac{1}{2} \delta ''((m_b v-q)^2-m_j^2) \mbox{.}
 \label{eq:deltas}
\eea
From the dispersion relation in the variable $v.q$ it can be 
seen \cite{man} that this replacement is only legitimate if one integrates 
over the whole neutrino phase space. This is no restriction as the 
$\tau$-neutrino is not detected.

As we are only interested in the double differential 
rate with respect to $q^2$ and $E_\tau $ we integrate over
$E_{\nu }$. Taking into account the $\delta$--function structure according
to (22) the integral over $E_\nu$ takes the following form
\begin{equation}
\int_{E_\nu^{min}(q^2,E_\tau)}^{E_\nu^{max}(q^2,E_\tau)}\mbox{d}\!E_\nu
  [f_1\delta(E_\nu-E_\nu^0)+f_2\delta'(E_\nu-E_\nu^0)+f_3\delta''(E_\nu
  -E_\nu^0)]\,,
\end{equation}
where
\begin{equation}
E_\nu^0 = \frac{m_b^2-m_j^2+q^2}{2m_b} - E_\tau\,.
\end{equation}
The limits of integration in (23) are as follows: i) for the values of
$(q^2,E_\tau)$ lying inside of region A of Fig.1 (for which the minimum
mass of the final hadronic state coincides with $M_{X_{min}}$, the mass of 
the lightest hadron state containing a quark $j$),
\begin{eqnarray}
E_\nu^{min} &=& \frac{q^2-m_\tau^2}{2(E_\tau+\sqrt{E_\tau^2-m_\tau^2})}\\
E_\nu^{max} &=& \frac{M_{H_b}^2-M_{X_{min}}^2+q^2}{2M_{H_b}}-E_\tau\,;
\end{eqnarray}
ii) for the values of $(q^2,E_\tau)$ inside of region B of Fig.1,
\begin{eqnarray}
E_\nu^{min} &=& \frac{q^2-m_\tau^2}{2(E_\tau+\sqrt{E_\tau^2-m_\tau^2})}\\
E_\nu^{max} &=& \frac{q^2-m_\tau^2}{2(E_\tau-\sqrt{E_\tau^2-m_\tau^2})}\,.
\end{eqnarray}

  The continuous line which separates the two regions A and B is given
by
\bea
q^2 = \frac{M_{H_b}(M_{H_b}^2-M_{X_{min}}^2+
   m_\tau^2-2M_{H_b}E_\tau)(E_\tau-\sqrt{E_\tau^2-m_\tau^2})
   +m_\tau^2M_{X_{min}}^2}{M_{H_b}^2+m_\tau^2-2M_{H_b}E_\tau}\,.
\nonumber\\
\eea

As in the massless lepton case, the actual particle masses are so that
always $E_\nu^0 < E_\nu^{max}$ as long as $(q^2,E_\tau)$ stay within region
A of the Dalitz plot. This is however no longer true for region B, where
this inequality has to be imposed explicitly. Its effect is to give an
upper bound on the mass of the final hadronic state (smaller than the
kinematically allowed one ($M_{H_b} - m_\tau$)). This is a new aspect which
was not present in the massless lepton case.

  The net result of all the restrictions imposed by the $\delta$ functions
and its derivatives is that the double differential decay rate (23) is
nonvanishing only within the partonic region delimited by the interrupted
line in Fig.1, determined in terms of quark masses. Its expression is
\begin{eqnarray}
&&
\frac{d\Gamma}{\Gamma_b dy d\hat q^2}= 
\nonumber\\&& \Theta (z_+)\Theta (-z_-)
\left( 12( - \hat q^4  + 2 \hat q^2 y 
                                + \hat q^2 \rho - \hat q^2 \eta
              - \hat q^2 - y^2  - y \rho + y \eta + y + \rho \eta - \eta)
           \right. 
\nonumber\\ &&
     +12 E_b (2 \hat q^4  - 2 \hat q^2 y - 2 \hat q^2 \rho + 2 \hat q^2 \eta
            + y \rho - y \eta + y - 2 \rho \eta)
\nonumber\\ &&     
    + 8 K_b ( - \hat q^4  + \hat q^2 \rho - \hat q^2 \eta + 2 \hat q^2 - 3 y
            + \rho \eta + 6 \eta) 
\nonumber\\ &&  \left.
    + 8 G_b (2 \hat q^4  - 2 \hat q^2 y - 2 \hat q^2 \rho + 2 \hat q^2 \eta
            - \hat q^2 + y \rho - y \eta - 2 y - 2 \rho \eta + 5 \eta)
     \right)
\nonumber\\ && 
-\delta(z_+)\left(
24 E_b \sigma_+  ( - \hat q^2 y - 2 \hat q^2 \sigma_+ + 2 \hat q^2
        + y^2  + 2 y \sigma_+ - y \eta - 2 y - 2 \sigma_+ \eta + 2 \eta)
           \right. 
\nonumber\\ &&
+ 4 K_b 
(\hat q^2 y^2  + 4 \hat q^2 y \sigma_+ + 4 \hat q^2 \sigma_+^2              
              - 20 \hat q^2 \sigma_+ - 4 \hat q^2 \eta - y^3  
             + y^2  \eta + 4 y \sigma_+^2       
              + 4 y \sigma_+ \eta 
\nonumber\\ &&  \hspace*{1.3cm} 
+ 12 y \sigma_+ + 4 y \eta + 4 \sigma_+^2  \eta
              -20 \sigma_+ \eta - 4 \eta^2 )
\nonumber\\ &&
 + 16 G_b  (\hat q^4  - \hat q^2 y \sigma_+ - \hat q^2 y             
              - 2 \hat q^2 \sigma_+^2  + \hat q^2 \sigma_+ 
- \hat q^2 + y^2  \sigma_+ + 2 y \sigma_+^2
              - y \sigma_+ \eta + y \sigma_+ 
\nonumber\\ &&  \hspace*{1.4cm} 
\left.+ y \eta 
- 2 \sigma_+^2  \eta - \sigma_+ \eta
              - \eta^2  + \eta)                    
     \right)
\nonumber\\ && 
+\delta(z_-)\left(
24 E_b \sigma_-  ( - \hat q^2 y - 2 \hat q^2 \sigma_- + 2 \hat q^2
        + y^2  + 2 y \sigma_- - y \eta - 2 y - 2 \sigma_- \eta + 2 \eta)
           \right. 
\nonumber\\ &&
+ 4 K_b 
(\hat q^2 y^2  + 4 \hat q^2 y \sigma_- + 4 \hat q^2 \sigma_-^2              
              - 20 \hat q^2 \sigma_- - 4 \hat q^2 \eta - y^3  
             + y^2  \eta + 4 y \sigma_-^2       
              + 4 y \sigma_- \eta 
\nonumber\\ &&  \hspace*{1.3cm} 
+ 12 y \sigma_- + 4 y \eta + 4 \sigma_-^2  \eta
              -20 \sigma_- \eta - 4 \eta^2 )
\nonumber\\ &&
 + 16 G_b  (\hat q^4  - \hat q^2 y \sigma_- - \hat q^2 y             
              - 2 \hat q^2 \sigma_-^2  + \hat q^2 \sigma_- 
- \hat q^2 + y^2  \sigma_- + 2 y \sigma_-^2
              - y \sigma_- \eta + y \sigma_- 
\nonumber\\ &&  \hspace*{1.4cm} 
\left.+ y \eta 
- 2 \sigma_-^2  \eta - \sigma_- \eta
              - \eta^2  + \eta)                    
     \right)
\nonumber\\ && 
+8 \delta'(z_+) \sigma_+ K_b\left(
  4 \hat q^4  - \hat q^2 y^2                 
    - 4 \hat q^2 y \sigma_+ - 4 \hat q^2 y - 4 \hat q^2 \sigma_+^2         
        + 4 \hat q^2 \eta + y^3  + 4 y^2  \sigma_+ \right.
\nonumber\\ &&  \hspace*{2.5cm} 
\left. - y^2  \eta + 4 y \sigma_+^2  
                   - 4 y \sigma_+ \eta - 4 \sigma_+^2  \eta  \right)
\nonumber\\ && 
-8 \delta'(z_-) \sigma_- K_b\left(
  4 \hat q^4  - \hat q^2 y^2                 
    - 4 \hat q^2 y \sigma_- - 4 \hat q^2 y - 4 \hat q^2 \sigma_-^2         
        + 4 \hat q^2 \eta + y^3  + 4 y^2  \sigma_- \right.
\nonumber\\ &&  \hspace*{2.5cm} 
\left. - y^2  \eta + 4 y \sigma_-^2  
                   - 4 y \sigma_- \eta - 4 \sigma_-^2  \eta  \right)
 \label{eq:dgaqy}
\end{eqnarray}
where 
\begin{eqnarray}
 z_\pm =1+\hat q^2 -\rho-y-2\sigma _\pm ,\qquad
     \sigma_\pm =(\hat q^2-\eta)\xi_\pm,\qquad\xi_\pm = \frac{1}
    {y\pm \sqrt{y^2-4\eta}}\,,
 \label{eq:sigma}
\end{eqnarray}
\begin{eqnarray}
 y&=& \frac{2 E_\tau}{m_b}, \hspace*{0.5cm}
 \hat q^2 =\frac{q^2}{m_b^2}, \hspace*{0.5cm}
 \rho = \frac{m_j^2}{m_b^2}, \hspace*{0.5cm}
 \eta = \frac{m_\tau^2}{m_b^2}
\end{eqnarray}
and 
\begin{eqnarray}
 \Gamma_b&=&\frac{\vert V_{bj} \vert ^2 G_F^2 m_b^5}{192 \pi ^3}
 \nonumber \mbox{.}
\end{eqnarray}

  In order to obtain the lepton spectrum, the double differential decay
rate (\ref{eq:dgaqy}) has to be integrated over $q^2$. The integration
limits are
\begin{eqnarray}
\hat q_{\pm} ^2 &=& 
\frac{(1-\rho + \eta -y)\left( y \pm \sqrt{y^2-4\eta }\right) 
 +2\rho \eta }{2(1+\eta -y)}
 \label{eq:limits}
\end{eqnarray}

The upper limit $\hat q^2_+$ corresponds to $z_+=0$ and the lower one
$\hat q^2_-$ to $z_-=0$.
  
The result is
\begin{eqnarray}
&& \frac{d\Gamma}{\Gamma_b dy}=
\nonumber\\ &&
\sqrt{y^2 - 4\eta} \left(
\left[ 2(-2y^2+3y\eta+3y-4\eta)
     - 6y\rho
     - \frac{6(-4\eta+y+y\eta)}{(1+\eta-y)^2}\rho^2 \right.\right.
\nonumber\\ &&
\left.\left.
     + \frac{-16\eta+6y+6\eta y-2y^2}{(1+\eta-y)^3}\rho^3 \right]\right.
\nonumber\\ &&
\left.
+ E_b \left[ 4(-y^2+3y-2\eta)
     + \frac{12(-6\eta+2\eta^2+y+3\eta y-\eta y^2)}{(1+\eta-y)^3}\rho^2
\right.\right.
\nonumber\\ &&
\left. \left.
     + \frac{4(20\eta-4\eta^2-6y-14\eta y+4y^2+4\eta y^2-y^3)}
            {(1+\eta-y)^4}\rho^3  \right] \right.
\nonumber\\ &&
\left.
+ K_b \left[ -\frac{4}{3}(-20\eta+9y+2y^2) \right.\right.
\nonumber\\ &&
\left.\left.
   - \frac{4(-20\eta+32\eta^2+4\eta^3+3y+2\eta y-25\eta^2 y+2y^2+4\eta y^2
         +2\eta^2 y^2-2y^3+\eta y^3)}{(1+\eta-y)^4}\rho^2 \right.\right.
\nonumber\\ &&
\left. \left.
   - \frac{4}{3(1+\eta-y)^5} (80\eta-104\eta^2+8\eta^3-18y-40\eta y
                             +50\eta^2 y+10y^2 \right.\right.
\nonumber\\ &&
\left.\left.          +22\eta y^2-8\eta^2 y^2-5y^3-5\eta y^3+y^4)
          \rho^3 \right]
+ G_b \left[ -\frac{4}{3}(-20\eta+15y+2y^2) \right.\right.
\nonumber\\ &&
\left.\left.
       -\frac{8(4\eta-3y-3\eta y+2y^2)}{(1+\eta-y)^2}\rho
       -\frac{4(12\eta+4\eta^2-3y-15\eta y+3y^2+2\eta y^2)}
         {(1+\eta-y)^3}\rho^2 \right.\right.
\nonumber\\ &&
\left. \left.
   -\frac{8(-20\eta+4\eta^2+6y+14\eta y-4y^2-4\eta y^2+y^3)}
         {3(1+\eta-y)^4}\rho^3 \right] \right)\,.
\end{eqnarray}

   The total decay rate can be obtained by integrating the spectrum
formula over the partonic range of values for $y$
\begin{equation}
2\sqrt{\eta} \leq y \leq 1-\rho+\eta\,,
\end{equation}
with the result
\begin{eqnarray}
\Gamma = \Gamma_b [f_0(\rho,\eta) + E_b f_E(\rho,\eta)
      + K_b f_K(\rho,\eta)  + G_b f_G(\rho,\eta)]\,.
\end{eqnarray}
Here
\begin{eqnarray}
f_0(\rho,\eta) &=& \sqrt{R}(1-7\eta-7\eta^2+\eta^3-7\rho+12\eta\rho
            -7\eta^2\rho-7\rho^2-7\eta\rho^2+\rho^3)\nonumber\\ & &
    -  24[\eta^2(1-\rho^2)\log\frac{1+\eta-\rho-\sqrt{R}}{2\sqrt{\eta}}
          + (\rho\leftrightarrow\eta)]\\
f_E(\rho,\eta) &=& \sqrt{R}(5-19\eta+5\eta^2-3\eta^3-19\rho-4\eta\rho
            +21\eta^2\rho+5\rho^2+21\eta\rho^2-3\rho^3)\nonumber\\ & &
    -  24[\eta^2(1+3\rho^2)\log\frac{1+\eta-\rho-\sqrt{R}}{2\sqrt{\eta}}
          + (\rho\leftrightarrow\eta)]\\
f_K(\rho,\eta) &=& 2\sqrt{R}(-3+13\eta+\eta^2+\eta^3+13\rho-4\eta\rho
                  -7\eta^2\rho+\rho^2-7\eta\rho^2+\rho^3)\nonumber\\ & &
    +  48[\eta^2(1+\rho^2)\log\frac{1+\eta-\rho-\sqrt{R}}{2\sqrt{\eta}}
          + (\rho\leftrightarrow\eta)]\\
f_G(\rho,\eta) &=& 2\sqrt{R}(-1+7\eta+7\eta^2-\eta^3+7\rho-12\eta\rho
                  +7\eta^2\rho+7\rho^2+7\eta\rho^2-\rho^3)\nonumber\\ & &
    +  48[\eta^2(1-\rho^2)\log\frac{1+\eta-\rho-\sqrt{R}}{2\sqrt{\eta}}
          + (\rho\leftrightarrow\eta)]
\end{eqnarray}
where
\begin{eqnarray}
R &=& (1+\eta-\rho)^2 - 4\eta\,.
\end{eqnarray}

   The limit of a massless final state quark $\rho \to 0$ requires a
special treatment because of the appearance of $\delta(y-1-\eta)$ and
$\delta'(y-1-\eta)$ functions in the lepton spectrum. They are due to
the fact that in this limit $z_+$ vanishes along the entire line
$y=1+\eta$. By considering $z_+$ as a function of $y$ at fixed $\hat q^2$,
one can write
\begin{eqnarray}
\delta(z_+) &=& \frac{1-\eta}{1-\hat q^2}\delta(y-1-\eta) + \cdots\\
\delta'(z_+) &=& \frac{(1-\eta)(1-\xi_+ y)}{(1-\hat q^2)[-1+\xi_+y+2\xi_+^2
   (\hat q^2-\eta)]}\delta'(y-1-\eta)+\cdots\,.
\end{eqnarray}
Inserting these relations into Eq.(30) and performing the
$\hat q^2$--integration between the limits
\begin{eqnarray}
\hat q^2_\pm = \frac{1}{2}\left(y\pm\sqrt{y^2-4\eta}\right)\,,
\end{eqnarray}
one obtains
\begin{eqnarray}
&& \frac{d\Gamma}{\Gamma_b dy}=\sqrt{y^2 - 4\eta} \left[
 2(-2y^2+3y\eta+3y-4\eta)\right.
\nonumber\\ &&
\left.
+ 4E_b (-y^2+3y-2\eta+\frac{1}{2}(1-\eta)^3\delta(y-1-\eta))\right.
\nonumber\\ &&\left.
-\frac{4}{3}K_b(-20\eta+9y+2y^2+(1-\eta)^2(5-2\eta)\delta(y-1-\eta))
\right.\nonumber\\ &&\left.
-\frac{4}{3}G_b(-20\eta+15y+2y^2-(1-\eta)^2(4-\eta)\delta(y-1-\eta))
 \right]\nonumber\\ &&
+8K_b(1-\eta)h(y,\eta)\delta'(y-1-\eta) \,.
\end{eqnarray}
Here $h(y,\eta)$ is defined by
\begin{eqnarray}
h(y,\eta) &=& \xi_+(1-\xi_+y)\int_{\hat q^2_-(y)}^{\hat q^2_+(y)}
   \mbox{d}\!\hat q^2\frac{(\hat q^2-\eta)}{(1-\hat q^2)[-1+\xi_+y+2\xi_+^2
   (\hat q^2-\eta)]}\times\nonumber\\& &
  \left(4 \hat q^4  - \hat q^2 y^2 - 4 \hat q^2 y \sigma_+
  - 4 \hat q^2 y - 4 \hat q^2 \sigma_+^2         
  + 4 \hat q^2 \eta + y^3
\right.\\ & & \left.
  + 4 y^2  \sigma_+
  - y^2  \eta + 4 y \sigma_+^2  
                   - 4 y \sigma_+ \eta - 4 \sigma_+^2  \eta  \right)
\nonumber
\end{eqnarray}
and satisfies
\begin{equation}
\lim_{y\to 1+\eta} \frac{\partial}{\partial y}h(y,\eta) = -\frac{2}{3}
(1-\eta)^2\,.
\end{equation}
An integration of the spectrum (45) over $y$ within the limits
$2\sqrt{\eta}\leq y\leq 1+\eta$ gives the total rate 
\begin{eqnarray}
\Gamma = \Gamma_b [g_0(\eta) + E_b g_E(\eta)
      + K_b g_K(\eta)  + G_b g_G(\eta)]
\end{eqnarray}
with
\begin{eqnarray}
g_0(\eta) &=& 1-8\eta+8\eta^3-\eta^4-12\eta^2\log\eta\\
g_E(\eta) &=& 5-24\eta+24\eta^2-8\eta^3+3\eta^4-12\eta^2\log\eta\\
g_K(\eta) &=& -6+32\eta-24\eta^2-2\eta^4+24\eta^2\log\eta\\
g_G(\eta) &=& -2+16\eta-16\eta^3+2\eta^4+24\eta^2\log\eta\,.
\end{eqnarray}
This is the same as what one would obtain by directly taking the limit  
$\rho\to 0$ in (36). In doing the $y$-integration of (45), the end-point
$y=1+\eta$ must be considered as included in the region of integration.

Before plotting the lepton energy spectrum we need to specify the values of 
the nonperturbative matrix elements $G_b$ and $K_b$. The matrix element 
$E_b$ is given by $E_b=G_b+K_b$. $G_b$ is related to 
the hyperfine splitting of degenerate heavy meson partners and is taken as  
$G_b=-0.0065$ as in \cite{man}. For $K_b$ we take $K_b=0.01$.
Further parameters are $m_b=4.8$ GeV and $m_c=1.39$ GeV. 
We shall plot the energy spectrum for mesonic b-decays into c and u.

Let us first discuss the decay spectrum (34)
for b $\to c$ plotted in Fig.3. The leading order contribution 
corresponding to free quark decay is larger than the $O(1/m_b^2)$ corrected 
spectrum over most of the $y$-range. The corrections become larger as the 
lepton energy increases. The corrected spectrum takes an upward turn to the 
high end of the spectrum. The OPE can, however, no longer be trusted in
local sense at the high end of the spectrum. As a rough estimate where
this happens we demand that the nonperturbative terms should not exceed 
30\% of the leading parton model result. Judging from the curve in Fig.3 this 
means that we trust the energy spectrum locally up to $y \lsim 0.98$. However, 
assuming duality, the integrated spectrum is expected to be given correctly 
even including the high end of the spectrum.

   The total inclusive decay rate $B\to X_c\tau\bar\nu_\tau$ is
\bea
 \Gamma/ \Gamma_b = 
(0.122 + 1.286 \cdot E_b - 1.409 \cdot K_b - 0.245 \cdot G_b)
= 0.122 - 0.008
\eea
for $m_\tau=1.78$ GeV,  $m_j= m_c=1.39$ GeV and $m_b=4.8$ GeV,
whereas for $m_\tau=0$ we would have had
\bea
 \Gamma_{m_\tau=0}/ \Gamma_b = (0.543 +3.361 \cdot E_b -3.904 
 \cdot K_b -1.086 \cdot G_b) = 0.543 - 0.020\,.\nonumber\\
\eea
The lepton mass effects appear thus to increase the procentual
contribution of the nonperturbative $1/m_b^2$ corrections in the total
result from --3.7\% to about --6.6\%.

Next turn to the tau spectrum for $b\to u$, which is plotted in Fig.4.
Also shown is the free quark decay spectrum for comparison. Again one
should trust the spectrum locally only for $y \lsim 0.98$. 

   The total decay rate is
\bea
 \Gamma/ \Gamma_b = 
(0.371 + 2.584 \cdot E_b - 2.954 \cdot K_b - 0.741 \cdot G_b)
= 0.371 - 0.016\,,
\eea
where the value $m_u=0$ has been taken.

While preparing this manuscript for publication we became aware of the 
paper of Koyrakh \cite{koy}, who has also considered lepton mass effects 
within the operator product expansion method. While we agree with his
final numerical results, we do not agree with all his formulas, which we
suspect to be marred with misprints.

\newpage

\section*{Figure Captions}
\begin{itemize}
\item[Fig.1] The Dalitz plot for the inclusive decay $B\to X_c\tau\bar
\nu_\tau$. The interrupted line shows the region of the phase space
which is populated in the parton model.
\item[Fig.2] The Dalitz plot for the inclusive decay $B\to X_u\tau\bar
\nu_\tau$. The interrupted line shows the region of the phase space
which is populated in the parton model.
\item[Fig.3] {$\tau$--energy spectrum for inclusive semileptonic 
bottom meson decays $b\to c$, normalized through division by
$\Gamma _b$. The continuous line is the spectrum including $O(1/m_b^2)$
corrections and the dotted line shows the leading parton model result.}
\item[Fig.4] {$\tau$--energy spectrum for inclusive semileptonic 
bottom meson decays $b\to u$, normalized through division by
$\Gamma _b$. The continuous line is the spectrum including $O(1/m_b^2)$
corrections and the dotted line shows the leading parton model result.}
\end{itemize}

\end{document}